\documentclass[
reprint,
 amsmath,amssymb,
 aps,
 showkeys,
pre,
]{revtex4-2}

\usepackage{graphicx}   
\usepackage{amsmath,amssymb,amsfonts} 
\usepackage{xcolor}     
\usepackage{booktabs}   
\usepackage{listings}   
\usepackage{subfig}

\newcounter{inlineenum}
\renewcommand{\theinlineenum}{\roman{inlineenum}}

\usepackage[english]{babel}
\makeatletter
\@namedef{l@en}{\@nameuse{l@english}}
\makeatother

\begin{document}

\title[Heterogeneous collectives]{Hydrodynamic interactions mask the true heterogeneity of a microscopic collective}

\author{Balagopal Nair}
\email{balagopalrnair1@gmail.com}
\affiliation{Department of Physics, IISER Pune, Dr Homi Bhabha Rd, Pune 411008, Maharashtra, India}

\author{Arshed Nabeel}
\email{arshed@iisc.ac.in}
\affiliation{Center for Ecological Sciences, IISc, Bangalore 560012, Karnataka, India}
\affiliation{IISc Mathematics Initiative, IISc, Bangalore 560012, Karnataka, India}

\author{Danny Raj M}
\email{danny@iitm.ac.in}
\affiliation{Applied Mechanics and Biomedical Engineering, IIT Madras, Chennai 600036, Tamil Nadu, India}

\begin{abstract}
    Coordinated movement and self-organisation of active self-driven agents is common in nature and is seen across different scales, from herds of animals to collective motion in bacteria. Often, these systems are heterogeneous in composition, with different agents having different intrinsic motilities. Inferring these intrinsic characteristics and quantifying the level of heterogeneity in a collective system is crucial to understanding the observed emergent phenomena. However, when interaction effects dominate, i.e. the observed movement of an agent is strongly influenced by its interacting neighbours, inferring the intrinsic characteristics of agents becomes a challenge. We consider a collective system of agents that undergo purely physical interactions like collisions and long-range hydrodynamic interactions, which resembles a system of microswimmers immersed in a fluid medium. 
    We incorporate heterogeneity into the system through variations in agent motility and examine how the perceived heterogeneity, inferred from measured speeds, depends on the strength of hydrodynamic interactions and the true intrinsic variability. 
    The interplay between short-range collisions, long-range hydrodynamic interactions, and intrinsic heterogeneity makes the inference problem non-trivial. When hydrodynamic effects dominate, true heterogeneity is effectively masked, making even a homogeneous collective appear heterogeneous. The competing effects of collisions, which slow agents down, and hydrodynamic interactions, which enhance their motion, further complicate reliable inference. Hydrodynamic interactions also modify collision angles, rendering them more isotropic. Overall, the findings show highlight experimentally measured properties of microscopic collectives may not accurately reflect their true characteristics.
\end{abstract}




\maketitle

\section{Introduction}\label{sec1}

Collective movement is a ubiquitous phenomenon, observed widely in natural as well as engineered systems, where individuals move and interact with each other to produce emergent, self-organized patterns~\cite{vicsek_collective_2012, ramaswamy_mechanics_2010}. Such collective movement can be observed across scales and systems, from cells~\cite{alert_physical_2020} and microbes~\cite{kearns_field_2010, beer_phase_2020} to animals~\cite{sumpter_principles_2006, couzin_self-organization_2003} and human crowds~\cite{helbing_social_1995,bacik_lane_2023}. The interactions between individuals in these systems could be along a spectrum of predominantly behavioural, like in animal groups and human crowds, to predominantly physical, like in microbial swarms or artificial systems. The broad question of how these local inter-individual interactions lead to ordered dynamics at the group level has fascinated biologists, physicists and engineers alike. 

Perhaps equally interesting is the inverse problem of inferring individual properties and interaction rules, based on observations of the collective dynamics of the group. Here, the goal is to start from the observed trajectories of individuals in a collective, and infer properties of the constituent individuals or reconstruct their interaction rules. Various approaches have been proposed for the inverse problem, across domains~\cite{zadeh_inferring_2024, bruckner_learning_2023, lukeman_inferring_2010, katz_inferring_2011, herbert-read_inferring_2011, calovi_disentangling_2018, nabeel_disentangling_2022}.

\begin{figure*}
    \centering
    \includegraphics[width=0.75\linewidth]{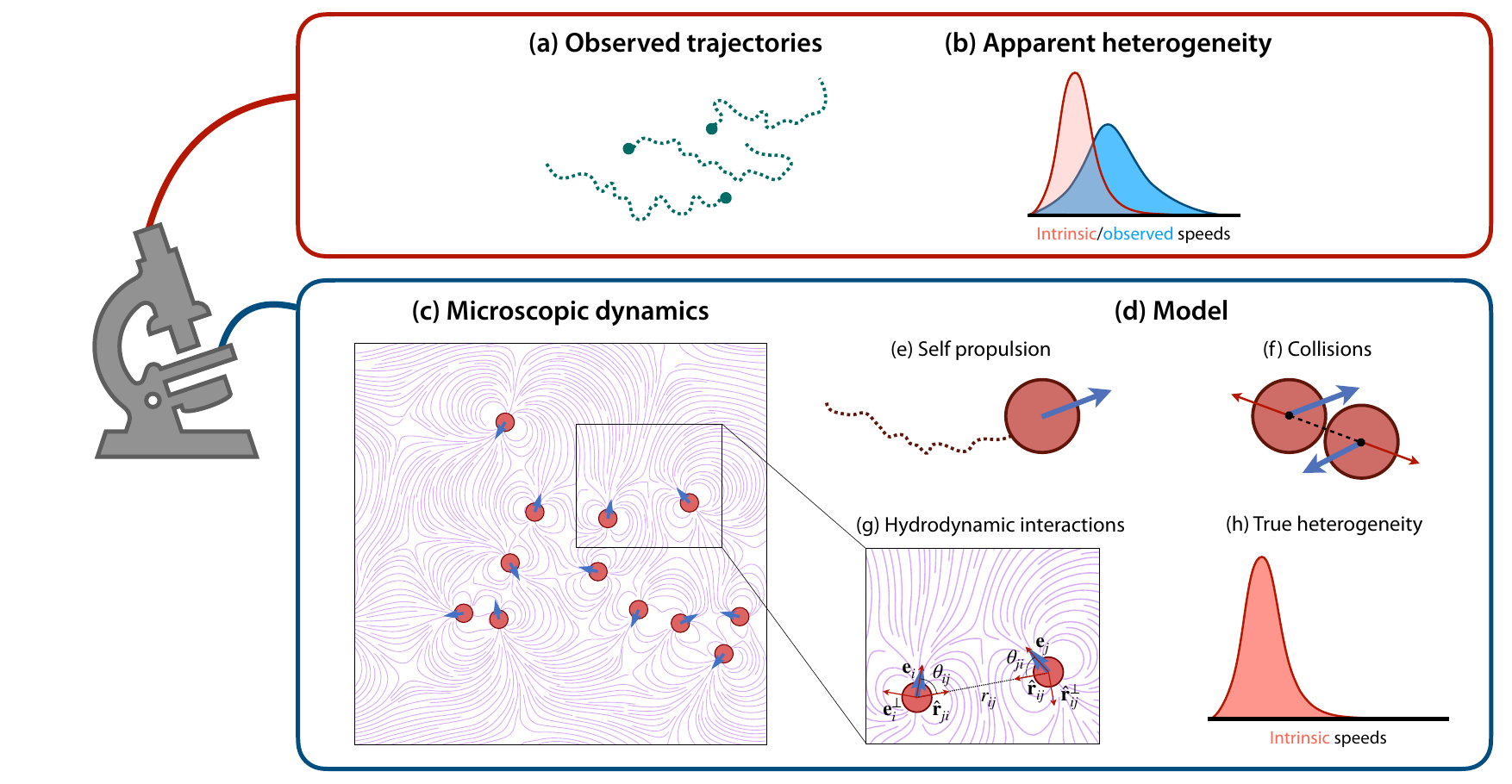}
    \caption{Schematic illustration of the problem: estimating heterogeneity from observations of a collective. (a, b) For an observer, the goal is to infer intrinsic properties of individuals by observing the collective dynamics. (a) In a typical setting, we have access to individual trajectory data. (b) We can estimate the level of apparent heterogeneity (for a property such as movement speed), shown in blue. This may be different from the true, intrinsic heterogeneity in the system (red). (c-f) We use a self-propelled particle model with hydrodynamic interactions for collective movement simulations. (e) Individuals self-propel with some angular noise with some intrinsic speed $v_{0_i}$ which can vary across individuals. The individuals interact through (f) soft collisions and through (g) far-field, flow-mediated hydrodynamic interactions. (h) The intrinsic speed of the individuals is sampled from a distribution.}
    \label{fig:schematic}
\end{figure*}

Individual heterogeneity is a key feature of most collective systems. Individuals forming the collective may be of different types, for e.g. in mixed species animal groups~\cite{goodale_mixed_2020, ward_cohesion_2018} or charged bidisperse colloids~\cite{reichhardt_stripes_2007}; or may vary along a continuum in some property such as speed~\cite{del_mar_delgado_importance_2018, jolles_role_2020}. Heterogeneity can be of key functional importance in many systems, such as ant colonies~\cite{fernandez-lopez_foraging_2025} and migrating cells~\cite{blanchard_devil_2019, Chiu2025}. In heterogeneous collectives, an important aspect of the inverse problem is to detect the presence of heterogeneity, and quantify the degree of heterogeneity when present~\cite{nabeel_disentangling_2022, tan_intrinsic_2024}. Although straightforward at first sight, this is a tricky inference problem, as heterogeneous collective systems can appear nearly homogeneous~\cite{nabeel_disentangling_2022}, and homogeneous systems can display signatures of heterogeneity~\cite{schumacher_semblance_2017}. \textcolor{black}{This is illustrated in \ref{fig:schematic}. A microscopic collective system may consist of individuals with some intrinsic heterogeneity in their properties such as movement speeds. Their actual dynamics in turn is affected by interactions among themselves and with the fluid medium in which they move in. When one tries to estimate the intrinsic properties (like speed) from the observed trajectories, the observed properties may appear more heterogeneous than the intrinsic properties, due to the effect of interactions.}

In this paper, we consider a model of collective movement consisting of individuals interacting through purely physical interactions, namely, soft collisions and far-field hydrodynamic interactions. This model captures, for instance, the dynamics of microswimmers like bacteria immersed in a fluid. Since the fluid-mediated hydrodynamic interactions are long-ranged, they can significantly affect the observed movement properties of the individuals. To study this effect systematically, we vary both the intrinsic heterogeneity (in terms of the movement speed) of the individuals and the strength of hydrodynamic interactions, and study how the interactions affect the observed heterogeneity of the system.
We find that hydrodynamic interactions can make a system appear heterogeneous, even when there is no intrinsic heterogeneity in the system. This apparent heterogeneity is observed to increase with the strength of the hydrodynamic interactions. 
Our study identifies conditions under which reliable inferences may be possible, as well as those where distinguishing apparent heterogeneity from true variability becomes nearly impossible.
We show that hydrodynamics modifies the microscopic collision profiles between agents, and these altered interactions give rise to the observed heterogeneity.
Taken together, these results shed new light into how inter-individual interactions in a collective system creates or alters apparent heterogeneity in collective system.

\section{The study}\label{sec2}

\subsection*{A collective movement model with hydrodynamic interactions}

\newcommand{\bv}{\mathbf{v}}
\newcommand{\be}{\mathbf{e}}
\newcommand{\br}{\mathbf{r}}
\newcommand{\bF}{\mathbf{F}}
\newcommand{\boldeta}{\boldsymbol{\eta}}
\newcommand{\voi}{v_{0_i}}
\newcommand{\bu}{\mathbf{u}}
\newcommand{\bU}{\mathbf{U}}

We consider a model consisting of active Brownian particles, interacting through soft collisions and hydrodynamic interactions induced through the fluid. Assuming a low Reynolds number regime, the movement of an individual $i$ can be described by the following (overdamped) equations of motion, similar to the one used in~\cite{filella_model_2018}:

\begin{align}
    \dot \br_i(t) &= v_{0_i} \, \be_i(t) + \bF_h(t) + \bF_r(t) \\ \label{eqn:rotation}
    \dot \theta_i(t) &= \Omega_i(t) + D \eta(t) \\ 
    \be_i(t) &= \textcolor{black}{\left( \cos \theta_i(t), \; \sin \theta_i(t) \right)} 
\end{align}

Here, $\br_i$ and $\be_i$ are respectively the position and orientation vectors of individual $i$, and $v_{0_i}$ is its intrinsic speed. $\bF_h$ and $\bF_r$ represent the \textcolor{black}{``forces"} on $i$ due to hydrodynamic interactions and volume-exclusion (repulsion) interactions respectively, and $\Omega_i(t)$ is the rotational component of the hydrodynamic force (see below) $\eta$ represents the noise fluctuations in individual movement. \textcolor{black}{We assume the overdamped limit, with appropriate units such that the mobility coefficient becomes unity.}

The repulsion interaction is modelled as a spring-like force that acts when the agents are closer than a threshold distance $R$, i.e.,

\begin{equation}
\mathbf{F}_r =
\begin{cases}
- \sum_{j \neq i} k(R - r_{ij}) \, \hat{\mathbf{r}}_{ij}, & \text{if } r_{ij} < R, \\
0, & \text{otherwise}.
\end{cases}
\label{eq:repulsion}
\end{equation}

Here, $\br_{ij} = \br_i - \br_j$, $r_{ij} = |\br_{ij}|$ and $\hat \br_{ij} = \br_{ij} / r_{ij}$. $k$ is the `spring constant' for the repulsive interaction.

\textcolor{black}{In addition, we also consider the effect of hydrodynamics on the movement of the particles. Here, we consider an incompressible, potential flow for the fluid, and only consider far-field terms, ignoring higher-order effects.}
Following the approach of Filella et al.~\cite{filella_model_2018}, the hydrodynamic interactions are modelled using far-field interactions, under a potential flow approximation. \textcolor{black}{Briefly, these equations for hydrodynamic interactions come from treating each agent as a dipole with a positive and negative vortex separated by some distance $L$, computing a complex potential for this dipole by taking a linear combination of the potentials of each vortex, taking the $L \to 0$ limit. The derivative of this potential gives us the complex velocity, which converted to the cartesian plane gives us these equations~\cite{filella_model_2018,Tchieu2012,Newton2005}}.

\begin{align}
    \bu_{ji} &= I_f \, \frac{v_{0_j}}{r^2_{ji}} \left( \cos \theta_{ji} \hat \br_{ij}^\perp + \sin \theta_{ij} \hat \br_{ij}\right) \\
    \bF_i &= \sum_{j \neq i} \bu_{ji}
\end{align}

\textcolor{black}{Here, $\theta_{ij}$ and $\theta_{ji}$ are the angles that $\be_i$ and $\be_j$ respectively make with the line joining the centres of $i$ and $j$. Vectors $\hat \br_{ij}$ and $\hat \br_{ji}$ are unit vectors as mentioned above, and $\hat \br_{ij}^\perp$ is a unit vector that is orthogonal (at angle $+ \pi/2$) to $\hat \br_{ij}$. These notations are illustrated in (\ref{fig:schematic})(g).} 
\textcolor{black}{The strength of the hydrodynamic interaction scales with the intrinsic speed of the agent $v_{0_j}$ and a tunable parameter $I_f$.}


In addition, the hydrodynamic interaction have a rotational component, given by,

\begin{align}
    \Omega_i = \sum_{j \neq i} \be_i \cdot \nabla \bu_{ji} \cdot \be_i^\perp
\end{align}

Finally, the heading angle $\theta_i$ is perturbed by Gaussian white noise $\eta$, with $\langle\eta(t) \eta(t')\rangle = \delta(t - t')$. The parameter $D$ denotes the noise strength. \textcolor{black}{These fluctuations account for fast timescale fluctuations in the movement directions due to various intrinsic or extrinsic factors, which are not explicitly modelled. Hence, the angular noise strength $D$ is kept as independent parameter. }

\paragraph*{Heterogeneity in the collective}
Our main goal is to study the effect of heterogeneity on the collective movement, and how inter-individual interactions affect the perceived heterogeneity in the system. \textcolor{black}{In the real world, heterogeneity amongst individuals in a collective can be through various factors: examples include phenotypic variability such as size variations in fish schools~\cite{jolles_role_2020}, behavioural heterogeneity such as in ant groups~\cite{fernandez-lopez_foraging_2025}, leader-follower heterogeneity in cell migration~\cite{vishwakarma_dynamic_2020}, and heterogeneity in motility properties in microswimmers due to physical variations or defects~\cite{Chiu2025,tufoni_flagellar_2024}. In most of these cases, the heterogeneity is manifested as variations in movement speed across individuals.}

\textcolor{black}{The goal is to identify the relationship between \textit{intrinsic heterogeneity}, i.e. heterogeneity in the physical properties and motilities of the individuals in the system,  and \textit{apparent heterogeneity}, i.e. heterogeneity in the observable properties, which can arise due to intrinsic heterogeneities, but can also be affected by interactions.} We introduce intrinsic heterogeneity in the system by varying the intrinsic speed $\voi$ across individuals. We sample $\voi$ from a log-normal distribution with $\mu=1$ and $\sigma$ ranging from 0 to 0.8, \textcolor{black}{and study how this reflects in the apparent heterogeneity, measured as their observed movement speeds. Note that our model is a variable speed model; collisions and hydrodynamic interactions can change the movement speed of agents and create apparent heterogeneity in the observed movement speeds, even when the intrinsic individual speeds are homogeneous.}

\paragraph*{Simulations}
We simulated our model in 2 dimensions with periodic boundaries. Simulations are done with 60 circular agents, of radius $R=0.1$, in a square domain of length $L=3$, with the inter-agent interaction strength set to $k = 50$ and a rotational diffusivity $D=1$. The mean intrinsic speed, as mentioned before is set to $1$. This speed-to-size ratio is typical for motile bacteria like E. coli (size $\sim$ 1-2 $\mu$m, speed $\sim$ 30 $\mu$m/s). \textcolor{black}{The strength of the far field interactions, $I_f$, is varied from $0$ to $0.04$, to study the effect of hydrodynamic interactions on the perceived heterogeneity. Recall that $I_f = 0$ corresponds to the absence of hydrodynamic forces, where the interactions are through collisions alone.} The agents are initialized with random initial conditions and positions, and the simulation is run for 1000 time-steps \textcolor{black}{(10 dimensionless time units), with an integration timestep $dt = 0.01$}. The simulation is initialized for a 100 time steps \textcolor{black}{(1 time unit)} to eliminate transient dynamics, after which data is collected.

\paragraph*{Observing the collective}
Since the heterogeneity in the system is introduced in the intrinsic speeds, and it is the speed of agents that is affected by the hydrodynamic interactions, we measure the heterogeneity of the system from the observed speeds of the agents. We vary $I_f$ across simulations to study how the hydrodynamic interactions affects the observed heterogeneity and measure the observed speed distributions to see how much it varies from the intrinsic speed distributions. The intrinsic speeds of the agents, as mentioned before, are in turn sampled from a log-normal distribution, with $\mu = 1$ and $\sigma$ varying from 0 (fully homogeneous) to 0.8.

\section{Results and discussion}\label{sec2}

\begin{figure*}
    \centering
    \includegraphics[width=0.75\linewidth]{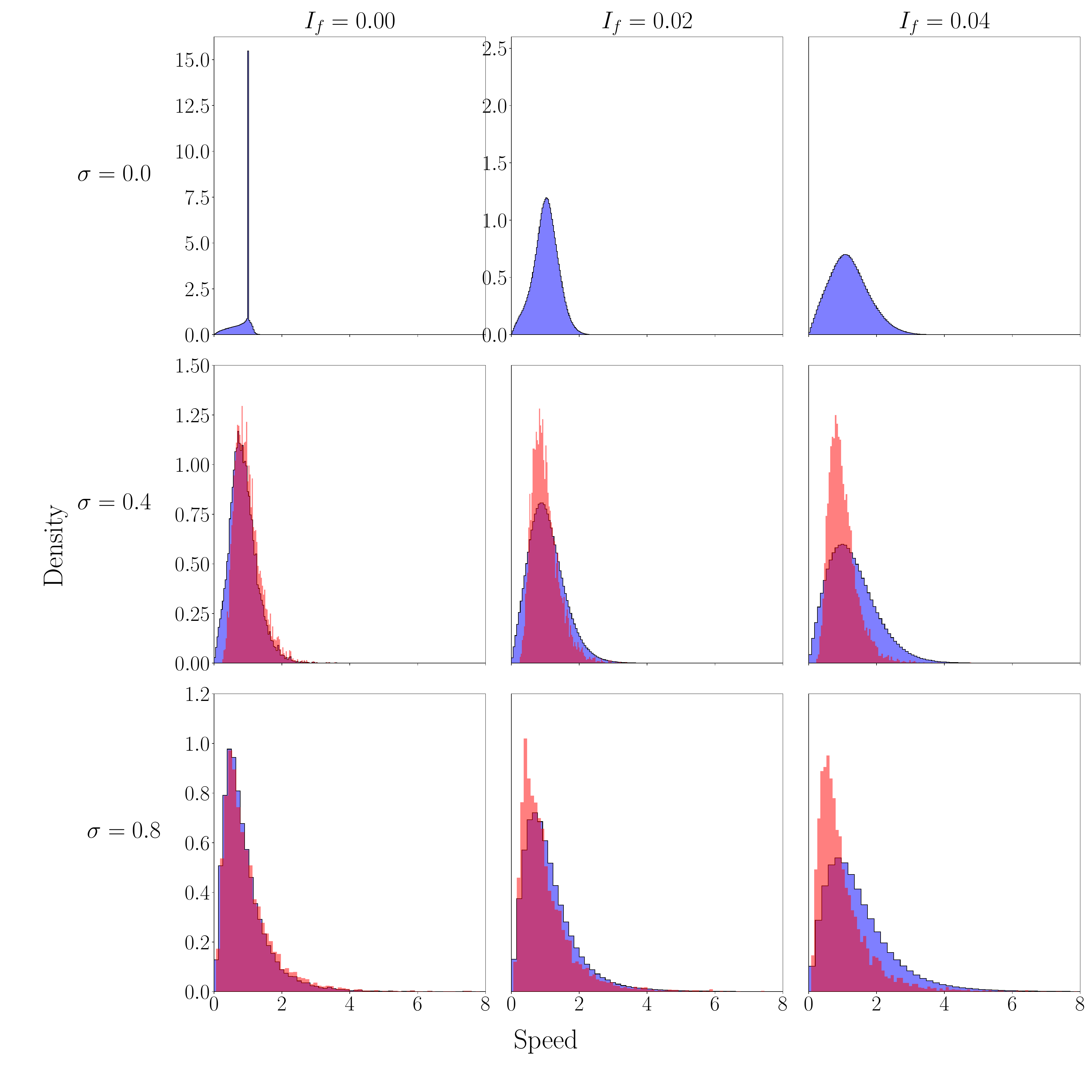}
    \caption{Distributions of observed speeds (in blue) taken over an ensemble of realisations for varying levels of heterogeneity $(\sigma)$ and hydrodynamic strength $(I_f)$. \textcolor{black}{The observed instantaneous values of the individual speeds over the entire simulation duration and multiple realisations, with no time averaging, are shown. At $\sigma = 0$, since the system is homogeneous only observed speeds are shown.} For values of $\sigma > 0$, the intrinsic speed distribution is also plotted in red. The figure illustrates how the observed speed distribution varies for different levels of heterogeneity and hydrodynamic strengths.}
    \label{fig:realisationscale}
    
\end{figure*}

\subsection{A homogeneous collective appears heterogeneous due to hydrodynamic interactions}\label{subsec2}
We first study the dynamics of a homogeneous collective where the agents that make up the system have identical intrinsic speeds (corresponds to the $\sigma=0$ case; first row in figure~\ref{fig:realisationscale}). \textcolor{black}{In the absence of any long-range hydrodynamic interactions (\textit{i.e.} when $I_f=0$), changes in the velocities of the active agents are only due to the collisions they experience as they move. Since the densities considered for our study are in the intermediate regime---between the dilute case where agents are isolated and the highly dense case where agents form jammed configurations---we find that agents experience collisions sparsely, which results in the collective appearing largely homogeneous. (\ref{fig:realisationscale}, $\sigma=0, \, I_f = 0$). Even when the intrinsic heterogeneity increases ($\sigma = 0.4$ and $\sigma = 0.8$), collisions alone have minimal effect on the observed heterogeneity of the system: the distribution of the intrinsic and observed speeds are quite similar to each other.} However, with increasing hydrodynamic interaction strength, $I_f>0$, we observe that the disturbance created in the fluid surrounding the agent leads to long-range interactions that results in large variations in the velocities of the agents making the collective appear heterogeneous. The observed spread has a long tail, indicating that hydrodynamic interactions constructively impact moving agents, allowing them to sample velocities much higher than their intrinsic velocities. This is in contrast to collision interactions which only reduce the maximum velocities that agents can sample every time they collide; \textit{i.e.}, slower agents speed up on colliding with a faster agent which slows down.

This result can be extended to the case where there is a discrete set of distinct intrinsic velocities among the agents. Say for instance, agents have either of two values of intrinsic speeds $v_0 = \{0.5, 1\}$. The distribution of the observed speeds in the presence of hydrodynamic interactions would be \textcolor{black}{similiar to what we saw in figure~\ref{fig:realisationscale} for the homogeneous case}, where the long-range interactions completely mask the distinct identities of the agents in the collective.

\subsection{Hydrodynamic interactions affect the perceived heterogeneity of the system}\label{subsec2}

\textcolor{black}{The previous section demonstrated that hydrodynamic interactions can significantly alter the speed heterogeneity of a system. In this section, we explore this in more detail from the perspective of an observer observing the collective dynamics, in terms of how the apparent heterogeneity relates to the true underlying heterogeneity of the system. We study this at two scales; at an \emph{ensemble} scale, and at an \emph{agent} scale.}

\textcolor{black}{Experiments with active microscopic collectives are usually one of two types. In the first type, we have a large number of relatively short-duration experiments. This facilitates calculating good population-level statistics of properties of interest, but we do not have information about truly long-term trajectories of any single agent. In the relatively rarer second type, we have longer duration experiments where only one or few individuals are tracked throughout the duration of the experiment. Here, we may or may not have a good handle of the population-level statistics of the collective, but know the long-term behaviour of individual agents. To understand the apparent heterogeneity in either of these cases, we use an ensemble scale and an agent scale analysis, corresponding to the first and second case respectively.
}



\paragraph*{Ensemble scale}
To generate the ensemble scale observations of the system, we treat all the data points (speed information of every agent at every time, in every realization) in our simulations as independent measurements, and we pool it together to construct the speed-distributions as observed in the various panels of figure~\ref{fig:realisationscale}.

\textcolor{black}{As observed for the homogeneous case, we find that for the $\sigma>0$ case}, that with increasing hydrodynamic strength ($I_f$), the collective appears more heterogeneous than what it originally was. Every panel in Figure~\ref{fig:realisationscale} has the distributions of the true heterogeneity (red) and the observed (blue). Hydrodynamic interactions alter the speeds of a focal agent depending on the dynamically changing local configuration of other agents around it.
We also observe an increase in the speeds of the agents indicated by the slight shifting of the mode of the distributions in figure~\ref{fig:realisationscale} towards the right with increasing $I_f>0$.
\textcolor{black}{This is not surprising, given that the hydrodynamic interaction strength is over and above the individual agent's self-propulsion. The momentum injected by self-propulsion is redistributed through the fluid and transmitted to neighbouring agents via hydrodynamic interactions, leading to an effective increase in their observed speeds. In contrast, when $I_f=0$, such momentum transfer through the fluid is absent and agents move solely under self-propulsion.}

\begin{figure}
    \centering

    \subfloat[]{%
        \includegraphics[width=0.85\linewidth]{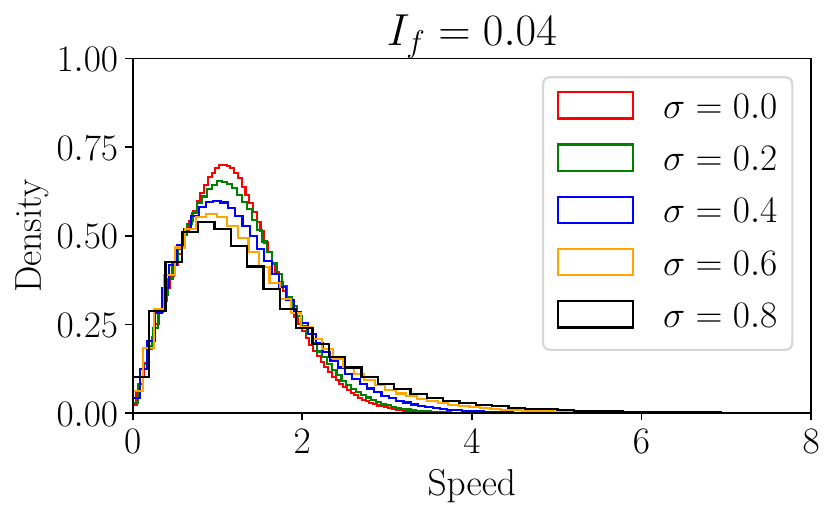}
        \label{fig:If004}
    }

    \vspace{0.5cm}

    \subfloat[]{%
        \includegraphics[width=0.85\linewidth]{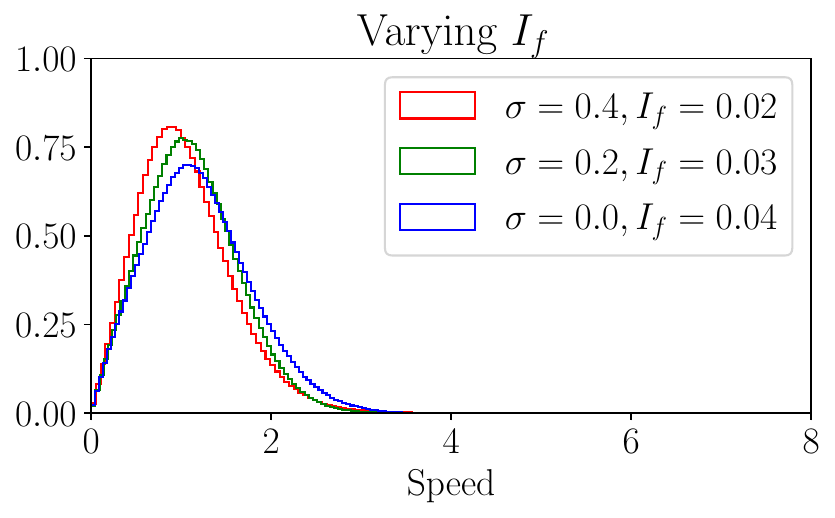}
        \label{fig:varyingIf}
    }

    \caption{Illustration of how the intrinsic speed of a system is masked by the presence of hydrodynamic interactions. 
    (a) Observed speed distributions of the system for high hydrodynamic strength $(I_f = 0.04)$ and varying heterogeneity $(\sigma)$. 
    (b) Observed speed distributions of systems along a diagonal direction in the parameter space --- from high $\sigma$ and low $I_f$ to low $\sigma$ and high $I_f$.}
    
    \label{fig:distinguishing}
\end{figure}

When the hydrodynamic interactions dominate (see figure~\ref{fig:distinguishing} A), we find that the true heterogeneity of the system is masked and the intrinsic speeds cannot be reliably inferred from the speed distributions.
Additionally, we find that it is difficult to distinguish collectives with low heterogeneity and large hydrodynamic effects from those with high heterogeneity and low hydrodynamic effects (see figure~\ref{fig:distinguishing} B).

Only when the intrinsic heterogeneity of the collective is large (ex: $\sigma \geq 0.6$) and the hydrodynamic strengths are low (ex: $I_f\leq0.01$), the observed heterogeneity reasonably matches the truth. 

\begin{figure}
    \centering
    \includegraphics[width=1\linewidth]{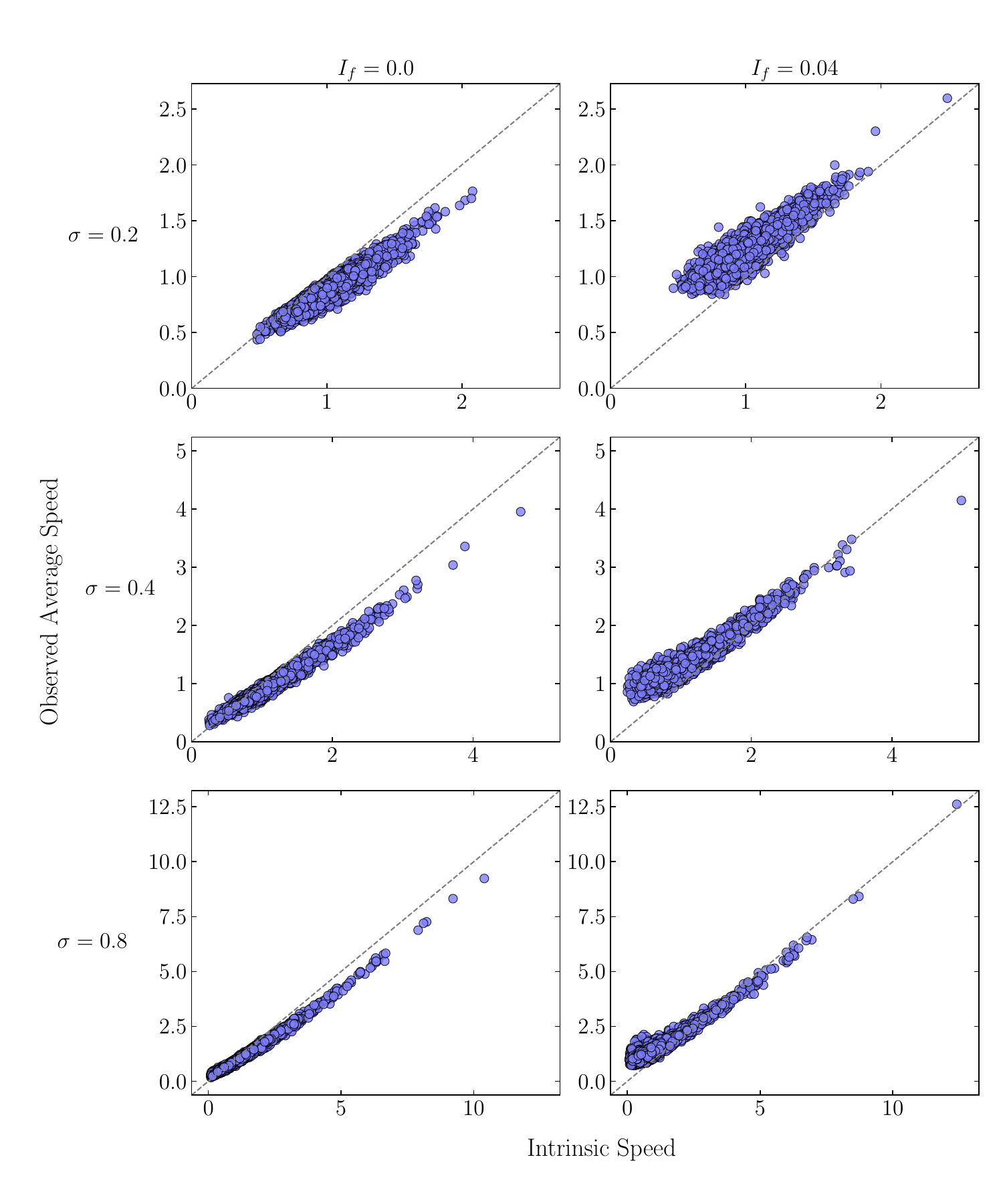}
    \caption{Plots of time-averaged speeds (over the entire simulation) of every individual agent in the collective across all independent realisations against their intrinsic speed for varying levels of heterogeneity $\sigma$ and hydrodynamic strength $I_f$. The dashed line corresponds to $x = y$. Each panel shows how average speeds of the agents change in comparison to the intrinsic speed for varying heterogeneity and hydrodynamic strengths.}
    \label{fig:agentscale}
\end{figure}

\paragraph*{Agent scale}
\textcolor{black}{To simulate the observations made in rare cases where agents maybe tracked throughout their lifetime, we take the time-averaged speeds of agents in the collective, over all independent realizations.} Here, averaging is done for the agents' speeds over the entire run of a simulation which is in contrast to the ensemble-scale observations where the speed at every time instant is treated as independent information.

Figure~\ref{fig:agentscale} shows the time-averaged observed speeds as a function of their intrinsic speeds. When the points lie on the $x=y$ line, the observed speeds match the true intrinsic speeds. Any departure is an over/under-estimation of the true speeds. In the absence of hydrodynamic interactions, we observe that observations match the truth for lower values of intrinsic speeds, while there is an underestimation of the speeds for agents with higher velocities. However, when hydrodynamic interactions are present, the trend is reversed. When the observed average speeds are lower, there is an overestimation of the intrinsic speeds of agents while higher observed speeds match reasonably with the truth. 
The trends observed are a consequence of the interplay between the collisions experienced by the agents and the hydrodynamic interactions.

\begin{figure}
    \centering
    \includegraphics[width=1\linewidth]{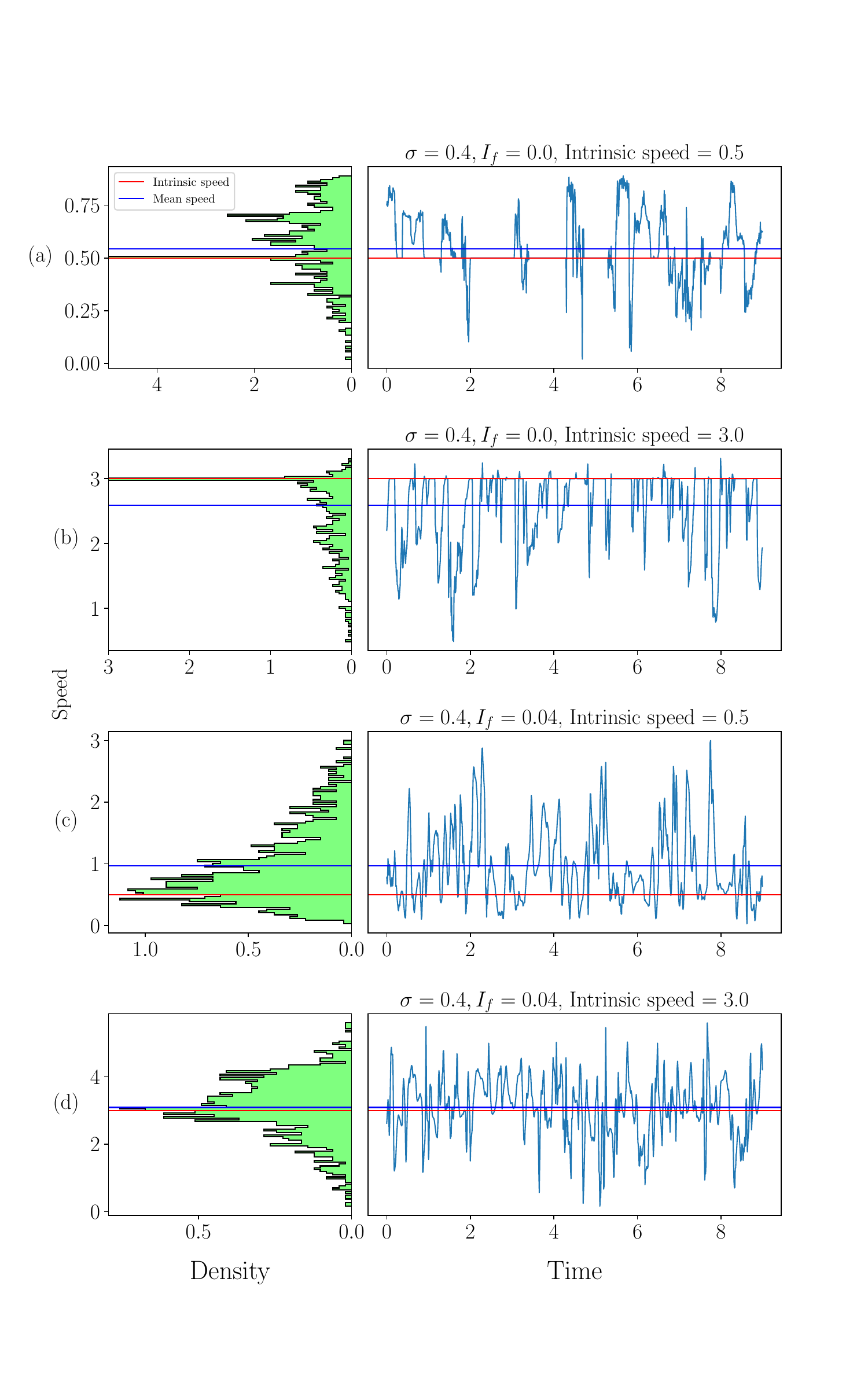}
    \caption{The time series and its distribution for the speed of a singular agent in a system with fixed heterogeneity $(\sigma = 0.4)$. We follow an agent with low intrinsic speed and high intrinsic speed, and collect observations for low (a,b) and high (c,d) hydrodynamic strengths $(I_f)$ respectively. }
    
    \label{fig:singleagent}
\end{figure}

\subsection{Hydrodynamics alters the collision-landscape between agents.}
To understand the role of collisions on the movement of an agent, we observe the time-series of the speeds of representative agents---one with a low intrinsic speed ($v_0$) and another with a high $v_0$. In the absence of hydrodynamics (figure~\ref{fig:singleagent} a, b), collisions have an asymmetric effect. Collisions always reduces the speed of a fast-moving agent, whereas for a slow-moving agent they can either decrease or increase its speed. 

\textcolor{black}{For instance, when two agents with the similar intrinsic (and similar instantaneous) speeds collide head-on, the agents are brought to a temporary standstill before the collision is resolved. In contrast, when one of the colliding agents is significantly slower, the collision initially slows down both the agents but subsequently reverses the direction of the slower agent. The slow agent then accelerates in its new direction as a consequence of the push it experiences from the fast moving one. Hence, a slow moving agent will likely speed up during collisions with faster moving agents.}
Therefore, the intrinsic speed of an agent with high $v_0$ is always under-estimated, while that of a slow moving agent is reliably inferred from average speed measurements.

However, in the presence of hydrodynamics (figure~\ref{fig:singleagent} c, d), the agents experience an additional overall push due to the disturbance flow of the fluid medium around them which results in higher overall speeds that compensate for that lost due to collisions. Hence, with increasing hydrodynamic strength $I_f$, the average speeds observed also increase, resulting in over-estimation of intrinsic speeds for slow-moving agents and reasonably reliable estimation for fast-moving agents.

To understand how the hydrodynamic interaction effects and heterogeneity affect the nature of collisions between agents we tabulate all the collision events in the collective. Here, a collision event begins when a pair of agents overlap upon approach, \textit{i.e.}, their inter-particle distance gets smaller than the sum of their radii ($r_{ij}<2R$), and it ends when the agent-pair no longer overlaps.

\begin{figure*}
    \centering
    \includegraphics[width=1\linewidth]{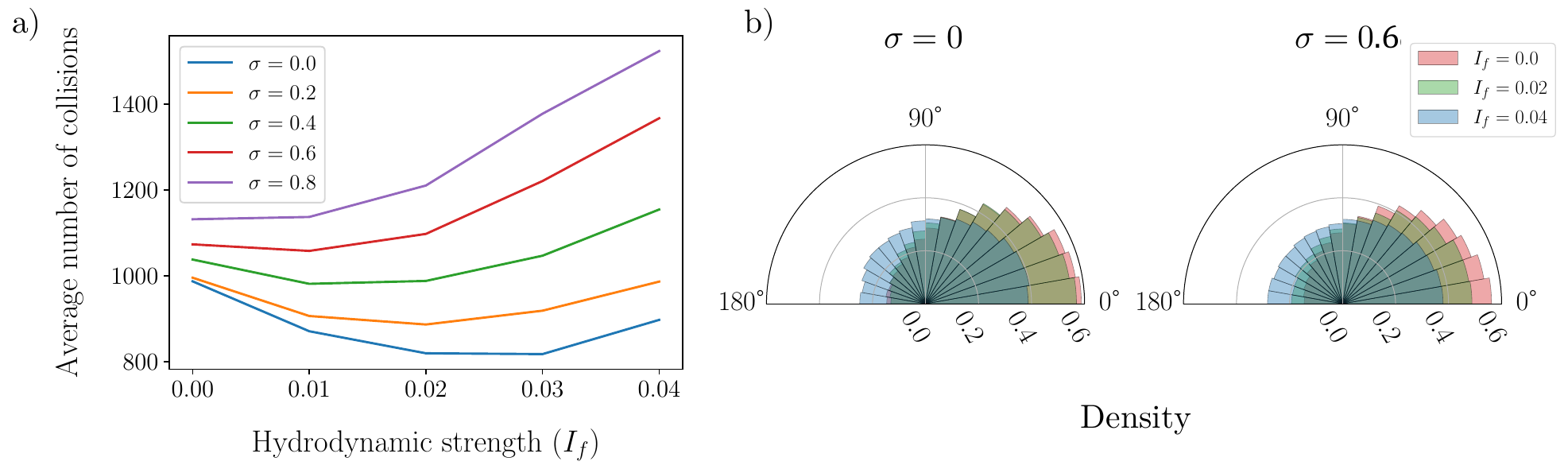}
    
    \caption{Effect of hydrodynamics on collision. (a) Average number of collisions in a system plotted as a function of the hydrodynamic strength $(I_f)$. In systems with higher heterogeneity $(\sigma)$, more collisions occur in general due to the greater variety in speeds of agents. As hydrodynamic strength increases, in systems with higher $\sigma$, collisions also increase, while in low $\sigma$ systems, it decreases first then increases. (b) Polar histograms that show the different angles at which collisions in the system take place. We observe that as hydrodynamic increases, the collisions become more isotropic for both low and high heterogeneity $(\sigma)$ cases.}
\label{fig:collisions}
\end{figure*}

We find that heterogeneity favours more collisions (see figure~\ref{fig:collisions} a). 
This is because agents with identical intrinsic speeds $v_{0}$ moving in similar directions cannot collide with one another. \textcolor{black}{However, when $v_0$ values are different they can cross paths due to the differential velocities thereby increasing the collisions between particles moving in similar directions.}

Collision dynamics are non-trivially affected by the hydrodynamic interactions between agents. The frequency of collisions varies non-monotonically with $I_f$ when $\sigma$ is small to none and increases monotonically for larger values of $\sigma$.
\textcolor{black}{To understand the above trends we need to understand how hydrodynamic effects affect the trajectory of the agents.}

\textcolor{black}{Hydrodynamic interactions not only alter the speed of the agents but also induce rotation} (see Eq~\ref{eqn:rotation}). Hence, we explore the collision angles between agent pairs in the collective. We compute the collision angle, which is the angle made between an agent's heading and the line joining the centres of the agent-pair just before the collision event (as shown in figure~\ref{fig:collisions}).

In the absence of hydrodynamics, we find that collisions are anisotropic, \textit{i.e.}, the angles are almost always `near' head-on ($<90^\circ$). However, in the presence of hydrodynamics, the collision profiles become `more' isotropic, \textit{i.e}, the agents sample higher collision angles ($>90^\circ$).

\textcolor{black}{Now when hydrodynamic interactions aren’t present, agents do not turn as they propel. In largely homogeneous systems, most agents move with similar speeds throughout the domain implying that collisions occur when agents primarily meet at angles close to $0^\circ$ (head-on). In the presence hydrodynamic effects, rotations in these agents increase, making head on collisions less frequent, and reducing the overall collision rate, which explains the initial dip we see in figure ~\ref{fig:collisions}\textcolor{black}{a}.}
\textcolor{black}{In addition to enhanced rotation, agent speeds are also modified by hydrodynamic interactions; however, we speculate that this effect alone is insufficient to significantly increase collisions among non-head-on pairs.}
\textcolor{black}{Upon further increasing $I_f$, spatial variations in the hydrodynamic fields leads to differences in the agent speeds, increasing the likelihood of collisions even in non-head-on configurations, and giving rise to the non-monotonic dependence as seen in figure~\ref{fig:collisions} a. For systems at higher heterogeneity, since speeds of agents are always different, and collisions occur over a wide range of angles. Increasing the hydrodynamic strength only amplifies the variation in the agents' speeds and turning rates, resulting in a larger number of collision events.}

Taken together, these observations show how the micro-level descriptions of collision dynamics are significantly affected by the long-range interactions between agents, which ultimately affect the reliability of the inferences made about the agents as seen in figures~\ref{fig:realisationscale}-~\ref{fig:agentscale}.

\section{Conclusion}
To summarize, we find that inferring the intrinsic characteristics and the true heterogeneity of a microscopic collective could be non-trivial because of the interplay between the heterogeneity, the short-term collisions and long-range hydrodynamic interactions in the system. 
Specifically, we find that hydrodynamic interactions can make an otherwise homogeneous collective appear heterogeneous. When hydrodynamic effects dominate, any true heterogeneity is effectively masked. Even when microscopic agents are tracked over longer time intervals, the competing effects of collisions that reduce the observed average speeds and hydrodynamic interactions that enhance them make accurate inference exceedingly difficult.
Hence, the characteristics of the system that we measure in an experiment with a microscopic collective may not represent the true properties highlighting the need for new methods to accurately infer them.

\textcolor{black}{
The simulations in this work were done in an intermediate density regime, where collisions are rare, and interactions are predominantly through the long-range, fluid-mediated interactions. Importantly, we have used potential-flow approximation, which we assume to be valid in this regime. In a high-density regime, while the potential-flow approximation may no longer hold, the interactions would be dominated by collisions, and it may be possible to ignore fluid interactions altogether. Conversely, in a low-density regime, the individuals can be very far from each other so any kind of interactions are rare and interaction-effects are negligible. Additionally, the potential flow approximation will not be valid near boundaries or inclusions, if the individuals are moving in a confined arena. Notwithstanding these limitations, we expect the broad point of this article--i.e. interactions can significantly alter apparent characteristics of a collective--to hold as a general principle.
}

In an earlier work, we showed that by computing a neighbourhood parameter, which accounts for the movement information of the near neighbourhood of an agent, we could disentangle the collective effects due to the interactions from the intrinsic characteristic of agents~\cite{nabeel_disentangling_2022}. The neighbourhood parameter captured how collisions either impeded or assisted the movement of an agent. However, the principle underlying this method was valid only for dense systems where collisions and close-range interactions dominated the collective dynamics, like the scenarios considered by Schumacher et al~\cite{schumacher_semblance_2017}. In this model, like mentioned above,  long-range interactions play the dominant role in the movement of the agents. \textcolor{black}{In these scenarios, approaches based on local information alone, such as the \textit{neighbourhood parameter} approach above, are insufficient to disentangle intrinsic motion from interaction effects.} Hence, formulations of the neighbourhood parameter, \textcolor{black}{explicitly accounting for long-range, non-local interactions}, are essential. \textcolor{black}{Coming up with a non-parametric and sufficiently general definition for such a non-local neighbourhood parameter is a challenging open problem for future work.}

\section*{Data availability statement}
No experimental data were collected in this study. The results are based on numerical simulations that can be reproduced using the model and parameters described in the paper. 

\section*{Acknowledgements}
AN thanks the MoE PhD Fellowship for the funding. And DRM gratefully acknowledges the support of the New Faculty Initiation Grant (NFIG) from IIT Madras, which facilitated the research presented in this manuscript.

\bibliography{sn-bibliography}

\end{document}